\documentstyle[12pt]{article}
\begin{document}

\begin{flushright}
Los Alamos preprint\\
LA-UR-93-3348\\
hep-ph/9309308\\
\end{flushright}

\baselineskip=22pt plus 0.2pt minus 0.2pt
\lineskip=22pt plus 0.2pt minus 0.2pt

\vspace{0.5in}

\begin{center}
\large
OBSERVABLE CONSEQUENCES OF A SCALAR BOSON\\
COUPLED ONLY TO NEUTRINOS\\

\vspace{0.50in}

G.\ J.\ Stephenson Jr.$^{\dagger}$ and T.\ Goldman\\ Los Alamos
National Laboratory\\ Los Alamos, NM 87545\\
\end{center}

\vspace{1.0in}

\begin{abstract}

We have examined the consequences of assuming the existence of a light
scalar boson, weakly coupled to neutrinos, and not coupled to any other
light fermions.  For a range of parameters, we find that this
hypothesis leads to the development of neutrino clusters which form in
the early Universe and which provide gravitational fluctuations on
scales small compared to a parsec (i.e., the scale of solar systems).
The existence of such clustering produces an effect which would appear
as a negative mass squared for the electron neutrino in Tritium beta
decay, without conflicting with other experiments.  The neutrino masses
arising in unified gauge theories would then be very much larger than
the masses extracted from experiments within the solar system.

\vspace{2.0in}

$^{\dagger}$Address after 1 Nov. 1993: Dept. of
Physics and Astronomy, University of New Mexico, Albuquerque, NM 87131

\end{abstract}

\setcounter{page}{0}

\pagebreak

In spite of the great success of the standard model for the description
of (electrically) charged fermions$^{1}$, there is still substantial
experimental ambiguity about the properties of neutrinos.  Experiments
designed to measure or limit neutrino masses appear to give negative
values for the mass squared$^{2-7}$; neutrinoless double beta decay,
which would confirm the expected Majorana nature of the neutrino$^{8}$,
has not been observed$^{9}$; and, if the neutrinos are Dirac particles,
nothing definite is known about the interactions of the right handed
neutrinos.  Theoretical extensions of the standard model to larger
gauge groups are, however, primarily constrained by interactions among
the charged fermions.

We have investigated the possibility that these and other experimental
data are evidence for an interaction mediated by a boson which couples
only to neutrinos and not to other light (electrically charged)
fermions$^{10}$.  Were this boson a vector, the need to cancel
anomalies arising from the coupling to one such vector and two Z$^0$'s
would require either coupling to charged fermions or the existence of
other new fermions designed for this purpose only.  We have, therefore,
studied the case of a scalar boson.  The theory of a system of
relativistic fermions interacting via scalar exchange is a special case
of Quantum Hadrodynamics$^{11}$ (QHD, developed for the study of
nuclear matter) some results of which we use here.  For clarity of
presentation, we shall concentrate on electron neutrinos, making
occasional reference to extensions to other flavors.

In this Letter we shall show that, for values of the scalar mass and
coupling constant which are compatible with known experimental data,
the coherent attraction arising from scalar exchange drives clustering
of neutrinos in the early Universe.  This causes them to decouple from
the general expansion at about the epoch of recombination and to
provide a source of gravitational fluctuations on a scale small
compared to a parsec yet large enough to influence stellar formation.
The existence of such clustering, persisting to the present epoch,
could provide sufficient neutrino capture events$^{2-6}$ to modify the
beta spectrum in Tritium beta decay at the end point and thus explain
the data.  A simple extension to include the interactions of muon
neutrinos with the background neutrinos, however, does not predict a
sufficient modification of the muon energy in pion decay to aacount for
the experimental observations$^{7}$.  Finally, we shall point out that
the effective masses which enter into the description of experiments
performed in the presence of such a background of clustered neutrinos
(i.\ e.\ within the solar system) are severely modified from the masses
that would be observed in the interstellar vacuum and which should
correspond to the mass parameters arising in gauge theories.
Derivations of the required formulas are deferred to a longer
paper$^{12}$.

Given a scalar boson, $\phi$, coupled to neutrinos, the interaction
Lagrangian density may be written in the form
$$
\cal{L} \rm_I = {\it g} \; \overline{\Psi_{\nu}} \; (cos \xi + i \gamma^5
sin \xi)\; \phi \; \Psi_{\nu} \eqno (1),
$$
where $g$ is the coupling constant and the angle $\xi$ fixes the
Lorentz properties of the coupling.   $\xi = 0$ yields a scalar
interaction, $\xi = \frac{\pi}{2}$ gives a pseudo-scalar coupling, and
angles in between lead to CP violation.  For this paper, we only
consider $\xi = 0$.

Following QHD, we express the effect of the medium on a single neutrino
of

\noindent momentum $k$ in terms of an effective mass,
$$
m^*_i(k) = m^o_{i} - \left(\frac{g_i}{2 \pi m_{\phi}}\right) \frac{m^*_i
(k)}{E^*_i (k)} \sum_j \left(\frac{g_j}{2 \pi m_{\phi}}\right) \frac{1}{\pi}
\int d^3k^{\prime} \frac{m^*_j (k^{\prime})}{E^*_j (k^{\prime})} F(k^{\prime})
\eqno (2)
$$
\noindent where $m^0$ is the vacuum value of the neutrino mass which
includes all field theoretic corrections except for the finite density
effect, $g$ is the coupling constant in (1), $m_{\phi}$ is the mass of
the scalar, $m^*(k)$ is the effective neutrino mass, $E^* =
\sqrt{m^{*2} + k^2}$ is the effective energy and $F(k)$ is the
Fermi-Dirac distribution function for the appropriate temperature. (We
use units with $\hbar = c = 1$.) The factors of  $m^*/E^*$ arise because
the effective mass corresponds to forward scattering which must
preserve helicity of both the subject neutrino and the background
neutrino, while the scalar coupling requires a chirality flip. We
differ slightly from Ref.\ 11 by defining the momentum-dependent mass
as the matrix element of the scalar self-energy operator at momentum
$k$, rather than the momentum-independent value at $k = 0$.  This
affords a clearer comparison with the covariant interaction energy as
discussed by MacRae and Riegert$^{13}$. The subscripts $i$ and $j$
refer to different flavors; for the bulk of the Letter we consider $i$
to refer to the electron flavor and suppress the sum.

The physical solutions of (2) are restricted to $0 < m^* < m^o$.
Writing
$$
A = \left(\frac{1}{m^o}\right) \frac{g^2}{4 \pi^3m^2_{\phi}} \int d^3
k^{\prime} \frac{m^* (k^{\prime})}{E^* (k^{\prime})} F (k^{\prime}) \eqno (3),
$$
we will show below that, for $A$ very large which is the case of
interest, $m^*(k)/ m^o < < 1$ for values of $k/m^o < A$.  This implies
that, for a large range of $k$, the neutrinos obey relativistic
kinematics and $E^* \simeq k$.  Hence neutrinos with $k < m^o$ do not
have enough energy to propagate freely outside the medium and are bound
to the system.  These remarks are illustrated in Fig.\ 1.

Consider now the evolution of the neutrino gas in the early Universe
from the point at which the weak interactions cease to be effective in
coupling neutrinos to other forms of matter.  As long as the
temperature is large compared to $m^o$, the difference between $m^*$
and $m^o$ is irrelevant and the neutrino gas is subject to the general
expansion.  When, however, the temperature drops to the order of $m^o$,
a large fraction of the neutrinos form a self-bound Fermi gas and will
nucleate into clusters whose size scale is set by $1/m_{\phi}$.  These
clusters, being self-bound, will retain their size while they separate
from each other due to the Universal expansion, although some
coalescence may occur.  Some of the most energetic neutrinos might well
escape from these clusters, providing a lower density of free relic
neutrinos than given by the standard model.  However, we estimate that
the energy loss rate due to bremsstrahlung of scalars from forward
scattering is sufficiently fast to cool much of the system, as will be
seen from the parameters below.

So far the discussion has been general and the scenario will develop at
some temperature and density, determined by $g$, $m^o$ and $m_{\phi}$.
To learn if the scenario can be realized for physically relevant
parameters, we need to be guided by data.

We begin with Tritium beta decay.  There are five modern experiments
$^{2-6}$, done by a variety of techniques and with various source
configurations, which study the end point of the electron spectrum to
search for evidence of non-zero electron anti-neutrino mass.  All
analyze the data under the assumption that the only extension of the
standard model is the existence of such a mass and that the signature
would be a suppression of the counting rate at and just below the end
point.  All five report a barely significant negative value for the
best fit to the square of the mass.  If real, this corresponds to an
increase in the counting rate in the vicinity of the end point.  In
fact, the Los Alamos group reported$^{6}$ that an equally good fit
could be obtained by assuming the existence of counts just at the end
point with a 10$^{-9}$ branching ratio.  This could occur by the
capture of massless relic neutrinos through the normal weak
interaction, providing the density is about $5 \times 10^{15}/cm^3$.
Such a density of cold neutrinos would correspond to a Fermi momentum
$(k_F)$ of $13 \, eV/c$; hence, as discussed below, the assignment of
all the counts to the endpoint is not justified.  A reanalysis, taking
into account the fact that $13 \, eV$ is not negligible on the scale of
the experiment and that such a density, if extended to anti-neutrinos,
provides additional distortion of the spectrum, is in progress$^{14}$.

Assuming that $A$ is large enough that $m^*$ is negligible for all $k <
k_F$, the argument goes as follows.  The density of final states is
calculated including the potential well that is represented by $m^*$
and gives, to observable accuracy, the same spectrum, with the same end
point $E_o$, as for free decay with zero neutrino mass.  However, the
well is filled with anti-neutrinos (right-handed Majorana neutrinos) so
Pauli blocking truncates the spectrum a distance $E^*= k_F$ below the
end point.  Since the well is also filled with neutrinos (left-handed
Majorana neutrinos) with a density proportional to $k^2$, they will be
captured and the beta spectrum will grow quadratically from $E_o$ up to
$E_o + k_F$ where it cuts off sharply.  The Kurie plot is shown in
Fig.\ 2.  Note that the interpretation of this differs from that
presented by Weinberg$^{15}$ and by Bergkvist$^{16}$ due to the
existence of the potential well.

To satisfy the condition that $m^*$ is negligible below $k_F$, we need
only require that $A>>1$.  To limit the parameters further, we must
consider other data.  One might be concerned that the existence of the
extra scalar degree of freedom, in thermal equilibrium with the
neutrino gas, would have an adverse effect on primordial
nucleosynthesis.  This question was studied by Kolb, Turner, and
Walker$^{17}$, who showed that, while a vector coupled either to
neutrinos or photons, or a scalar coupled to photons, would have
serious effects, the case at hand corresponds to adding an additional
half family of neutrinos which is not definitely ruled out by standard
analyses.  Furthermore, the efficacy of $\mu$- and $\tau$-neutrinos may
well be affected by this interaction, depending on the details of the
coupling to other flavors.  This question requires further study.
However, since the scalar does couple left- and right-handed
chiralities, if the neutrinos were Dirac particles, one would expect
twice the number of light neutrino species.  This would conventionally
cause difficulties with the calculated primordial abundances of nuclear
species.  Furthermore, the right-handed neutrinos and left-handed
anti-neutrinos would not interact with matter in a supernova and would
provide too efficient a cooling mechanism.  These considerations add
further credence to the assumption that the neutrinos are Majorana
particles.

The Lagrangian density (1) does lead to the annihilation of pairs of
neutrinos (with the same chirality) into pairs of scalars.  In order
that background neutrinos have an effect on Tritium beta decay, they
must persist until the present epoch.  This puts a very severe
constraint on the size of the coupling.  Writing
$\stackrel{\sim}{\alpha} = g^2/ 4 \pi$, and averaging over a zero
temperature Fermi gas, we obtain the mean annihilation rate per
neutrino, given by
$$
<\omega> = \frac{3}{8} \stackrel{\sim}{\alpha} ^2 k_F
\left(ln(\frac{k_F}{m_{\phi}}) - \frac{3}{8}\right) \eqno (4).
$$
Requiring  $<\omega>$ to be less than $10^{-18}/s$ constrains
$\stackrel{\sim}{\alpha} \; < 10^{-18}$ upon iterating with $m_{\phi}$
as determined below.(The functional form of this rate is generic for
all two body processes involving the scalars or neutrinos.)

A further constraint comes from the non-observation of neutrinoless
double beta decay.  From Tritium, we expect $m^o$ to be of the order of
$13 \, eV/c^2$ and, from the arguments above, that the neutrino is a
Majorana particle.  Neutrinoless double beta decay can arise from the
exchange of virtual Majorana neutrinos between the affected quarks in a
nucleus with the matrix element proportional to the effective
mass$^{18}$.  Of course, the effective mass  must be evaluated at
momenta appropriate to the size of the nucleus, or several hundred
$MeV/c$.  To safely suppress the zero-neutrino mode, $m^*$ should be
kept negligible at values of $k/m^o$ up to $10^8$ which demands $A >
10^9$, as we now show.

Writing $m_{\phi}/m^o = \mu$, $y = m^*/m^o$, and $x = k/m^o$, we may
rewrite (2) as
$$
y = 1 - \frac{\stackrel{\sim}{\alpha}}{\pi^2 \mu^2}\;
\frac{y}{\sqrt{x^2 + y^2}} \int d^3 x^{\prime}
\frac{y^{\prime}}{\sqrt{x^{\prime 2} + y^{\prime 2}}} F (x^{\prime}) =
1 - \frac{y}{\sqrt{x^2 + y^2}} A \eqno(5), $$
or, solving for $x^2$,
$$
x^2 = y^2 \left[\frac{A^2}{(1 - y)^2} - 1 \right] \simeq
\left(\frac{y}{1-y}\right)^2 A^2 \eqno (6).
$$
Thus to ensure $y$ does not approach 1 for $x < 10^8$, we must
have $A > 10^9$.

We now are in a position to bound the value of $m_{\phi}$.  With $A >
10^9$, $y = x/A$, to an extremely good approximation over all of the
range of integration supported by $F(x)$.  When the temperature equals
$m^o$, (3) may be written
$$
A \cong \frac{4 \tilde{\alpha}}{\pi \mu^2} \int^{\infty}_o x^2 d x \cdot
\frac{1}{\sqrt{A^2 + 1}} \frac{1}{(e^x + 1)} \eqno (7).
$$
This gives
\vspace*{-0.20in}

$$
\frac{\stackrel{\sim}{\alpha}}{\mu^2} \cong \frac{\pi A^2}{6 \zeta (3)}
\eqno (8)
$$
where $\zeta$ is the Riemann zeta function, $\zeta(3)$ = 1.20206.  From
the bound on $A$,
\vspace*{-0.20in}
$$
\frac{\stackrel{\sim}{\alpha}}{\mu^2} > 10^{18} \eqno (9).
$$

Thus, $\mu < 10^{-18}$, or $m_{\phi} < 10^{-17} \, eV/c^2$.  This
implies a range greater than $10^{12} \, cm$, consistent with the
notion that the neutrino clustering can provide seeds for the formation
of stars.  Clusters with large density and small extent (less than a
parsec) will behave as mass points (as do stars) for considerations of
galaxy formation and rotation curves.  The fraction of neutrinos that
is not in clusters may well play a role in the required dark matter
$^{19}$.

If there is a large density of relic neutrinos in the vicinity of the
Sun (and, of course, of the Earth), the scattering, by the background,
of neutrinos which are observed in experiments must not be too large,
providing further constraints on the parameters.  In the chirality
basis, the total neutrino-neutrino scattering cross sections are given
by\\

\vspace*{-0.55in}

\begin{eqnarray*}
\sigma_{LL} = \sigma_{RR} = \frac{\pi \stackrel{\sim}{\alpha} ^2}{s}
\left[ 4 + \frac{2(2m^2 - m^2_{\phi})^2}{m^2_{\phi}(s-4m^2 +
m^2_{\phi})}\hspace*{1.00in} \right. \nonumber \\ \left. +  4
\left\{\frac{(2m^2 - m^2_{\phi})}{(s - 4m^2)} - \frac{(2m^4 +
m^2_{\phi} (s - 4m^2) + m^4_{\phi})}{(s - 4m^2)(s - 4m^2 +
2m^2_{\phi})}\right\} ln \left(\frac{s - 4m^2 +
m^2_{\phi}}{m^2_{\phi}}\right)\right] \end{eqnarray*}\vspace*{-.65in}\\
\hspace*{5.50in} (10)\\

\noindent and

\begin{eqnarray*}
\sigma_{LR} = \frac{\pi \stackrel{\sim}{\alpha} ^2}{s} \left[ 1 + \frac{(2m^2 -
m^2_{\phi})^2}{m^2_{\phi} (s - 4m^2 +
m^2_{\phi})}\hspace*{0.40in}\right.\nonumber \\
\left. + \frac{2(2m^2 - m^2_{\phi})}{(s - 4m^2)} ln \left(\frac{s - 4m^2 +
m^2_{\phi}}{m^2_{\phi}}\right)\right]
\end{eqnarray*}\vspace*{-.60in}\\
\hspace*{5.50in} (11),\\

\noindent where $m$ is the appropriate neutrino mass and $s$ is the
total c.\ m.\ energy squared.

Those neutrinos which have been observed experimentally have laboratory
energies in excess of $1 \, MeV$, so the dependence on $s$ allows the
possibility of both the strong effects needed for large $A$ and a
sufficient mean free path to observe reactor or accelerator neutrinos
over kilometers, and solar neutrinos over an Astronomical Unit.  The
worst case is presumably given by the observation of neutrinos from
supernova SN1987a.  To obtain the strongest constraint, we use the most
conservative assumption.

Assume that all the neutrinos created in the early Universe survive to
the present epoch and are collected in clusters around the stars,
proportional to the baryon number associated with the stellar system.
Since the ratio of neutrinos to baryons is essentially the same as that
of photons to baryons $(10^9)$ the number of neutrinos associated with
the solar system would be $10^{66}$.  At a uniform density of $5 \times
10^{15}/cm^3$, they would fill a sphere of radius $3.6 \times 10^{16}
\, cm$, or about .01 parsec.  To achieve a mean free path greater than
$4 \times 10^{16} \, cm$, the cross section must be less than $5 \times
10^{-33}cm^2$.   The neutrinos observed from SN1987a had an energy of
about $10 \, MeV$ and the effective energy of the relic neutrinos is
about $10 \, eV$, so the average $s$ is $10 ^8 eV^2$.  The dominant
pieces of equations (10) and (11) then give
\vspace*{-0.20in}
$$
\sigma_{tot} \simeq \frac{5 \pi \stackrel{\sim}{\alpha}^2}{s} \eqno (12),
$$
which easily satisfies the cross-section bound if the annihilation rate
bound discussed above has been satisfied.

We note that, under these assumptions, the total energy of the
neutrinosphere approaches 10 solar mass equivalents.  We expect such a
mass concentration to be efficient at driving baryonic clustering.
Even if only a fraction of the neutrinos cluster, which eases the
cross-section constraint, the effect on baryon clustering should still
be strong.

If the predominant cluster size favors the formation of smaller baryon
agglomerations than a typical solar system, i.e., -``Jupiters'', then
much of the baryonic matter in the Universe will not be visible, and
the nucleosynthesis bound on the baryonic contribution to the universal
deceleration parameter may be saturated.$^{20}$ In that case, the
neutrino contribution to the universal energy density being an order of
magnitude larger from our estimate, the deceleration parameter may well
equal unity on large scales.

We should return here to the questions of annihilation and cooling. The
structure of the annihilation cross-section is similar to (12). We used
that together with the flux of essentially massless neutrinos in the
cluster to make the estimate (4) for the annihilation rate. The
cooling  depends on bremsstrahlung of scalars and their ability to
subsequently escape the neutrinosphere without rescattering.  Again,
the scattering rate of scalars on neutrinos is negligible because the
cross-section is similar to (12). Only coherent scattering, which
dominates the bremsstrahlung process, can be different.

In making the cooling rate estimate, we use a Weiszacker-Williams
approximation to estimate the probability of $\phi$-bremsstrahlung as
$P \sim (g^2/8 \pi^{2}) \; ln \; (E_{max}/E_{min})$, per neutrino
scattering.  The neutrinos bound in a cluster of size $1/m_{\phi}$ form
a coherent potential for which a naive estimate of the neutrino
scattering cross section would exceed the unitarity bound.  We
therefore use a geometrical cross section and take the flux to be that
given by a density of order the local density with each neutrino moving
relativistically.  The energy loss rate is then given by
$$
dE/dt \sim \frac{1}{\pi m^2_{\phi}} \cdot P \cdot \rm{flux} \cdot <
\rm{energy \; loss \; per \; scatter}> \eqno(13)
$$
The momentum transfer involved must be less than or of order $m_{\phi}$
for coherence to obtain over the entire cluster. For forward inelastic
scattering, one may estimate straightforwardly that the neutrino-scalar
pair mass in the c. m. of the cluster and a neutrino (and so, the
maximum energy of the scalar) is limited to $A m_{\phi}$. Using this,
we estimate that the energy lost to $\phi$ emission approaches the
binding energy of a neutrino essentially instantaneously. The decoupled
scalars then red-shift the energy away in the general expansion of the
Universe. The eventual result will be that a significant fraction of
the primordial neutrinos will form cold, self-bound Fermi gas clusters
of dimensions comparable to the original clusters.

The scalar boson under consideration is, in many respects, like the
Majoron proposed many years ago$^{21}$.  Limits on
$\stackrel{\sim}{\alpha}$ can be obtained from the bremsstrahlung of
the $\phi$ in elementary processes, such as $K \rightarrow \mu + \nu$,
or a study of the two electron spectrum in double beta decay, as in
analyses for Majorons.$^{22,23}$  The strongest reading of
these limits requires that $\stackrel{\sim}{\alpha} < 10^{-12}$, which
again is not constraining.

A constraint on the entire scenario, rather than on certain parameters,
may eventually be obtained from solar system dynamics.  For a cold,
self bound relativistic Fermi gas, the average energy per particle is
$3 k_Fc/4$.  Assuming $k_F = 10 \, eV/c$ and a density of $5 \times
10^{15}/cm^3$, one obtains an energy density of $4 \times
10^{16}eV/cm^3 = 7 \times 10^{-17}g/cm^3$.  The orbit of Jupiter is
$7.8 \times 10^{13}cm$, so the extra mass interior to the orbit is $1.3
\times 10^{26}~g = 10^{-7}M_{\odot}$.  To the best of our knowledge,
this is not ruled out by existing data. Future studies of the motion of
interplanetary probes might be able to determine the presence of the
neutrinosphere.

Finally, a comment on the muon neutrino is in order.  Note that the
parameters for the electron neutrino do not constrain the specific
coupling of the scalar to other flavors.  For example, if the vacuum
masses were proportional to the $g_i$, the effective masses would be in
the ratio of the vacuum masses.  Nonetheless, generally, the effect of
an attractive well on the muon neutrino is such that a neutrino emitted
in pion decay must have a larger momentum than it would outside the
neutrinosphere.  To balance momentum and conserve energy, the muon
momentum will be increased, with a concomitant increase in the muon
energy. However, independent of how large the vacuum mass value is,
this can not lead to a negative effective value of the square of the
neutrino mass$^{12}$, but can only reduce it to a negligibly small
value.  Thus the scenario described here does not account for the
results of Ref. 7.  Large vacuum masses for the non-electron neutrino
flavors would, of course, cause severe difficulty for our understanding
of the present expansion of the Universe unless those neutrinos
annihilate into scalars or convert into electron neutrinos sufficiently
rapidly, which could place lower limits on their coupling to scalars.
While such extensions are very interesting, and could ultimately
demonstrate the feasibility, or lack thereof, of this entire scenario,
they are beyond the scope of the present Letter.

In summary, we have shown that the assumption of a very light scalar
boson, coupled to neutrinos and to no other light fermions, can be
accommodated within existing experimental data.  The existence of such
a scalar can have serious implications for the evolution of matter
concentrations in the early Universe since it provides fluctuations on
a scale much smaller than those observed by COBE$^{24}$ (which
presumably drive the large scale structure).  These include the
possibility that neutrino clustering seeds stellar formation and so
influences the stellar mass distribution.  Furthermore, the clustering
of neutrinos driven by such a scalar interaction can provide an
explanation of the negative mass-squared found in modern Tritium beta
decay experiments.

A disturbing consequence for model builders is the fact that
experiments performed within the solar system will yield values of the
effective neutrino masses, not the vacuum masses that must arise from
an eventual unified theory.  In particular, it should be noted that the
$13 \, eV/c^2$ vacuum value of the electron neutrino mass discussed
above in no way obviates currently popular interpretations of the
observed value of the solar neutrino flux.

Taking the Tritium experiments as a datum, the clustering would be
manifest at a temperature of about $13 \, eV$, or near the time of
electron-ion recombination.  Accommodating the lack of neutrinoless
double beta decay leads to a minimum range of the scalar interaction
which is a reasonable fraction of an Astronomical Unit and allows a
range of the order of, or larger than, planetary orbits.  This
suggests that the primordial neutrino distribution could evolve into
droplets with a scale commensurate with solar systems, which would then
provide a seed mechanism for the formation of stars at all times from
recombination to the present epoch.

This work was performed under the auspices of the United States
Department of Energy.  We want to thank our colleagues at Los Alamos
and at the University of Melbourne, especially Bruce McKellar, for many
long and helpful discussions.

\pagebreak
\begin{center}
REFERENCES\\
\end{center}

\noindent $^{\dagger}$Address after 1 Nov. 1993: Dept. of
Physics and Astronomy, University of New Mexico, Albuquerque, NM 87131

\begin{enumerate}

\item T.\ Goldman in Adv.\ Nucl.\ Phys.\ (J.\ N.\ Negele, and
Erich Vogt, eds.\ ), Vol.\ {\bf 18}, p.\ 315  (1987) Plenum Press, NY.

\item Ch.\ Weinheimer, M.\ Przyrembel, H.\ Backe, H.\ Barth,
J.\ Bonn, B.\ Degen, Th.\ Edling, H.\ Fischer, L.\ Fleischmann,
J.\ U.\ Grooss, R.\ Haid, A.\ Hermanni, G.\ Kube, P.\ Leiderer,
Th.\ Loeken, A.\ Molz, R.\ B.\ Moore, A.\ Osipowicz, E.\ W.\ Otten,
A.\ Ricard, M.\ Schrader, and M.\ Steininger, Phys.\ Lett.\  {\bf
B300}, 210 (1993).

\item E.\ Holzschuh, M.\ Fritschi, and W.\ Kundig,
Phys.\ Lett.\  {\bf B287}, 381 (1992).

\item H.\ Kawami, S.\ Kato, T.\ Ohshima, S.\ Shibata, K.\ Ukai,
N.\ Morikawa, N.\ Nogowa, K.\ Haga, T.\ Nagafuchi, M.\ Shigeta,
Y.\ Fukushima and T.\ Taniguchi, Phys.\ Lett.\  {\bf B256}, 105
(1991).

\item W.\ Stoeffl, Bull.\ Am.\ Phys.\ Soc.\ Ser ll {\bf 37},
1286 (1992).

\item R.\ G.\ H.\ Robertson, T.\ J.\ Bowles, G.\ J.\ Stephenson, Jr.,
D.\ L.\ Wark, J.\ F.\ Wilkerson and D. A. Knapp, Phys.\ Rev.\ Lett.\
{\bf 67}, 957 (1991).

\item R.\ Abela, M.\ Daum, G.\ H.\ Eaton, R.\ Frosch,
B.\ Jost, P.\-R.\ Kettle, and E.\ Steiner, Phys.\ Lett.\ {\bf 146B},
431 (1984); B.\ Jeckelmann, T.\ Nakada, W.\ Beer, G.\ de Chambrier,
O.\ Elsenhans, K.\ L.\ Giovanetti, P.\ F.\ A.\ Goudsmit, H.\ J.\ Leisi,
O.\ Piller, and W.\ Schwitz, Phys.\ Rev.\ Lett.\ {\bf56}, 1444 (1986).

\item H.\ Primakoff and S.\ P.\ Rosen,
Ann.\ Rev.\ Nucl.\ Part.\ Sci.\ {\bf 31}, 195 (1981); and references
therein.

\item M.\ K.\ Moe, U.\ C.\ Irvine preprint, UCI-Neutrino 93-1;
T.\ Bernatowicz, J.\ Brannon, R.\ Brazzle, R.\ Cowsik, C.\ Hohenberg,
and F.\ Pobseh, Phys.\ Rev.\ Lett.\ {\bf 69}, 2341 (1992).

\item If, as is argued in the text, the neutrinos are Majorana
fermions, $\Gamma = i \gamma^o \gamma^2$ will achieve this result.

\item Brian D.\ Serot and John Dirk Walecka in
Adv.\ Nucl.\ Phys.\ (J.\ N.\ Negele, and Erich Vogt, eds.), Vol.\ {\bf
16}, p.\ 1  (1986) Plenum Press, NY.

\item G.\ J.\ Stephenson Jr.\ and T.\ Goldman, in
preparation.

\item K.\ I.\ MacRae and R.\ J.\ Riegert, Nucl.\ Phys.\ {\bf
B244}, 513 (1984).

\item R.\ G.\ H.\ Robertson, G.\ J.\ Stephenson Jr.,
J.\ F.\ Wilkerson, and D.\ A.\ Knapp, in preparation.

\item Steven Weinberg, Phys.\ Rev.\ {\bf 128}, 1457 (1962).

\item Karl-Erik Bergkvist, Nucl.\ Phys.\ {\bf B39}, 317 (1972).

\item Edward W.\ Kolb, Michael S.\ Turner, and Terrance
P.\ Walker, Phys.\ Rev.\ {\bf D34}, 2197 (1986).

\item W.\ C.\ Haxton and G.\ J.\ Stephenson Jr., in Progress in
Particle and Nuclear Physics (Sir Denys Wilkinson, ed.), Vol.\ 12, 409
(Pergamon Press, New York, 1984), and references therein.

\item For a discussion of the effect of non-interacting
neutrinos on rotation curves, and a guide to earlier literature, see
John P.\ Ralston and Lesley L.\ Smith, Ap.\ J.\ {\bf 367}, 54 (1991).

\item Peter J. Quinn, private communication.

\item Y.\ Chikashige, R.\ N.\ Mohapatra, and
R.\ D.\ Peccei, Phys.\ Lett.\ {\bf 98B}, 265 (1981); G.\ B.\ Gelmini
and M.\ Roncadelli, Phys.\ Lett.\ {\bf 99B}, 411 (1981);
H.\ M.\ Georgi, S.\ L.\ Gashow, and S.\ Nussimov, Nucl.\ Phys.\ {\bf
B193}, 297 (1981).

\item V.\ Barger, W.\ Y.\ Keung, and S.\ Pakvasa,
Phys.\ Rev.\ {\bf D25}, 907 (1982); T.\ Goldman, Edward W.\ Kolb, and
G.\ J.\ Stephenson, Jr.\ , Phys.\ Rev.\ {\bf D26}, 2503 (1982).

\item A.\ Piepke et al., U.\ of Heidelberg preprint, 1993;
J.\ C.\ Vuilleumier, J.\ Busto, J.\ Farine, V.\ J\"{o}rgens,
L.\ W.\ Mitchell, M.\ Treichel, J.\ L.\ Vuilleumier, H.\ T.\ Wong,
F.\ Boehm, P.\ Fisher, H.\ E.\ Henrikson, D.\ A.\ Imel, M.\ Z.\ Igbal,
B.\ M.\ O'Callaghan-Hay, J.\ Thomas, and K.\ Gabathuler,
Phys.\ Rev.\ {\bf D48}, 1009 (1993).

\item G.\ F.\ Smoot et al., Ap.\ J.\ {\bf 396}, L1 (1992).

\end{enumerate}

\vspace{2.0in}

\begin{center}
Figure Captions
\end{center}

Figure 1. Ratio of E$^*$/m$^0$ and m$^*$/m$^0$ as a function of
k/m$^0$.  Note the logarithmic scale for k/m$^0$, and that E$^*$/m$^0$
goes through 1 for k/m$^0 <$ 1.\\

Figure 2.  Modified Kurie plot, showing the square root of the counting
rate versus the energy of the electron.  The cut off is due to the
Pauli principle suppression by the filled anti-neutrino well; the
additional counts beyond the end point are due to the capture of
neutrinos bound in the well.

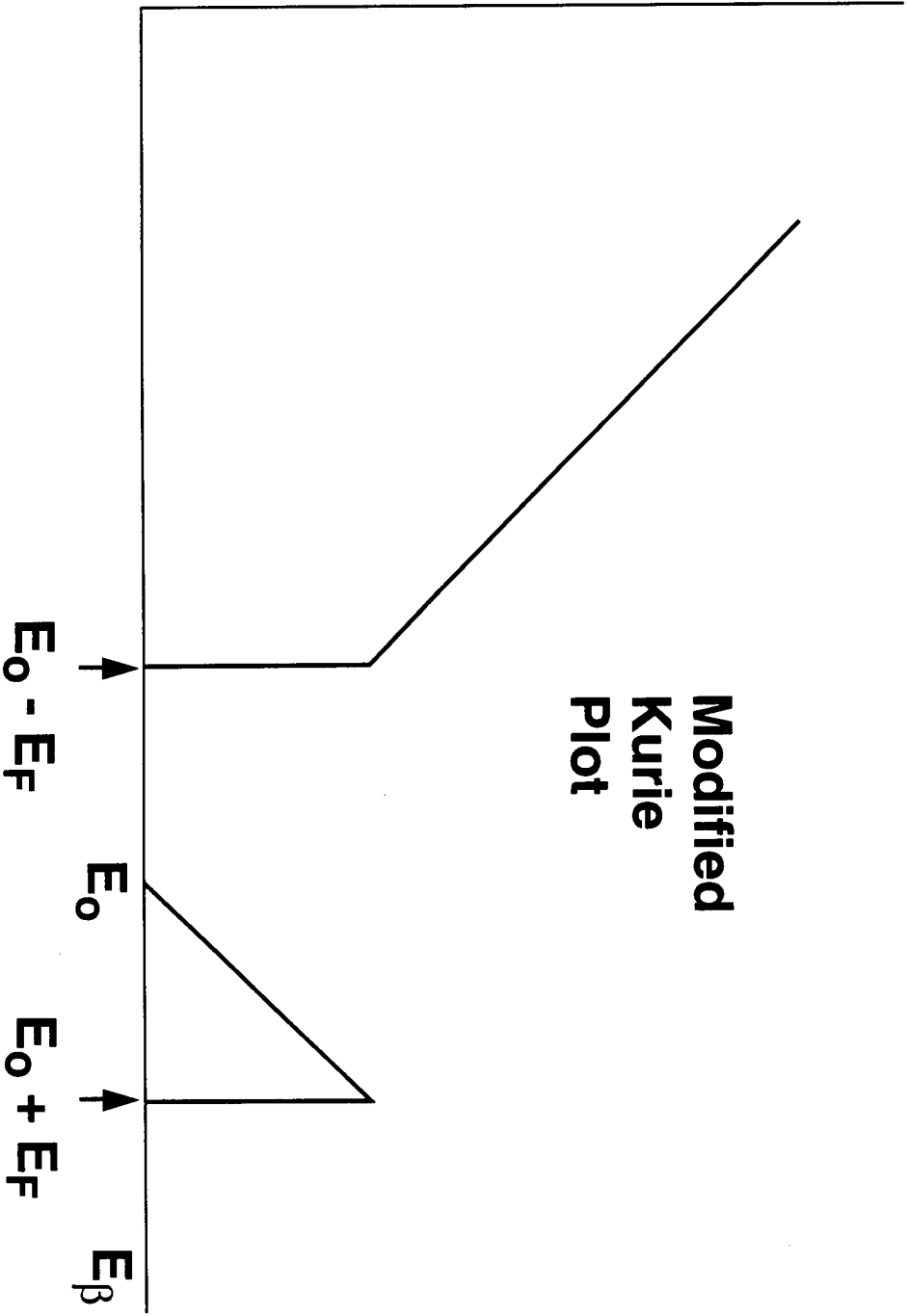
\end{document}